\documentclass[preprint,pteplogo]{ptephy_v2}

\preprintnumber{XXXX-XXXX} 
\usepackage{hyperref}

\usepackage{graphics} 
\usepackage{url} 

\begin{document}

\title{Feasibility demonstration of a high-momentum, high-purity muon beamline at the J-PARC Hadron Experimental Facility}


\author{Kotaro~Shirotori}
\affil[1]{Research Center for Nuclear Physics (RCNP), The University of Osaka, 10-1 Mihogaoka, Ibaraki, Osaka 567-0047, Japan \email{sirotori@rcnp.osaka-u.ac.jp}}

\author[1]{Takaya~Akaishi}

\author[2]{Kazuya~Aoki}
\affil[2]{Institute of Particle and Nuclear Studies (IPNS), High Energy Accelerator Research Organization (KEK), Tsukuba 305-0801, Japan}

\author[3]{Wen-Chen~Chang}
\affil[3]{Institute of Physics, Academia Sinica, Taipei 11529, Taiwan}

\author[2]{Ryotaro~Honda}

\author[4]{Yusuke~Hori}
\affil[4]{Department of Physics, Graduate school of Science, Kyoto University, Kyoto 606-8502, Japan}

\author[1]{Takatsugu~Ishikawa}

\author[1]{Nobuyuki~Kobayashi}

\author[2]{Ruri~Kurasaki}

\author[2]{Che-Sheng~Lin}

\author[2]{Yuhei~Morino}

\author[4]{Megumi~Naruki}

\author[1]{Hiroyuki~Noumi}

\author[2]{Kyoichiro~Ozawa}

\author[2]{Shinya~Sawada}

\author[1]{Ken~Suzuki}

\author[2]{Hitoshi~Takahashi}

\author[5]{Taiga~Toda}
\affil[5]{Department of Physics, Graduate School of Science, The University of Osaka, Toyonaka 560-0043, Japan}

\author{Natsuki~Tomida}
\affil[6]{Center for Science Adventure and Collaborative Research Advancement (SACRA), Graduate school of Science, Kyoto University, Kyoto 606-8502, Japan}

\begin{abstract}%

We have established and characterized a novel high-purity muon beam delivery scheme, 
designated as $\mu20$, at the high-momentum hadron beamline (B-line) of the J-PARC Hadron Experimental Facility. 
This scheme delivers tertiary muons produced via the in-flight backward decays of secondary pions in the straight section of the beamline, 
inherently suppressing parent hadron contamination. 
In a systematic performance evaluation, the muon beam achieved a standalone intensity of $1.7 \times 10^3~{\rm particles/spill}$ 
with a spatial spread (standard deviation, $\sigma$) of approximately $15~{\rm mm}$ at the downstream detector position. 
A systematic attenuation analysis utilizing an iron absorber 
verified a high muon purity of $98.0 \pm 0.6\%$ for the $3~{\rm GeV}/c$ muon beam mode. 
Crucially, this work marks the first successful feasibility study of a high-momentum and high-purity muon beamline operation 
at an experimental facility in Japan.

\end{abstract}

\subjectindex{G2, G20, H14}

\maketitle

\section{Introduction}

High-momentum (in-flight) muon beams within the multi-${\rm GeV}/c$ regime provide unique and complementary opportunities 
for the transmission radiography of large-scale infrastructure and dense geological structures, 
whereas high-intensity, low-energy muon beams have been extensively developed worldwide for precision particle physics and material sciences. 
Conventional muon radiography utilizes passive cosmic-ray muons; 
however, its experimental efficiency is fundamentally limited by the low intensity and steep spectral suppression. 
The integrated cosmic muon flux at ground level is approximately $1~{\rm particle/cm^2/min}$, 
and the differential energy spectrum exhibits a steep power-law suppression in the high-energy region. 
Consequently, this statistical limitation intrinsically caps the spatial resolution and necessitates prolonged data-acquisition periods for massive absorbers. 
In contrast, muon beams generated at high-energy accelerator facilities 
can provide a probe with fully controlled energy, high intensity, and adjustable collimation. 
These superior beam properties enable high-contrast, high-resolution radiographic imaging, 
facilitating the non-destructive visualization of internal density contrasts for critical applications 
such as identifying internal steel bars within thick concrete and inspecting the interior of heavily shielded radioactive waste casks.

According to the multiple Coulomb scattering theory~\cite{PDG}, 
the standard deviation of the scattering angle distribution ($\theta_{0}$) of a muon passing through a material is described as
\begin{equation}
\theta_{0} = \frac{13.6~{\rm MeV}}{\beta c p} \sqrt{\frac{L}{X_0}} \left[1 + 0.038 \ln \left( \frac{L}{X_0 \beta^2} \right) \right],
\end{equation}
where $L$ and $X_0$ are the path length and the radiation length of the material, respectively. 
Rearranging this relationship highlights the material dependence:
\begin{equation}
X_0 \propto L \left(\frac{13.6~{\rm MeV}}{\beta c p \theta_{0}}\right)^2.
\end{equation}
Since the velocity of a muon greater than $1~{\rm GeV}/c$ is essentially $\beta \sim 1$, 
$\theta_{0}$ is inversely proportional to the muon momentum ($p$) and depends on the radiation length $X_0$, 
which is a quantity intrinsic to the material. 
Because the radiation length $X_0$ is a characteristic property of each material, 
measuring the precise incoming and outgoing muon trajectories 
allows us to reconstruct the internal structure and material distribution of massive objects.

To realize such a multi-${\rm GeV}/c$ muon probe, 
we conducted the J-PARC T106 experiment~\cite{t106, t106proc} 
to extract a high-momentum and high-purity muon beam 
at the high-momentum beamline (B-line) of the J-PARC Hadron Experimental Facility~\cite{HDEF}. 
The present muon measurements provide fundamental data to evaluate the feasibility of performing muon radiography. 
Notably, this work marks the first successful demonstration of the feasibility of operating a high-momentum and high-purity muon beamline 
at an experimental facility in Japan.

\section{High-momentum beamline in the $\mu20$ mode} \label{sec:beamline_design}

\subsection{$\pi20$ beamline mode} \label{subsec:pi20_beamline}

\begin{figure}[t]
\centering
\includegraphics[width=16cm,clip]{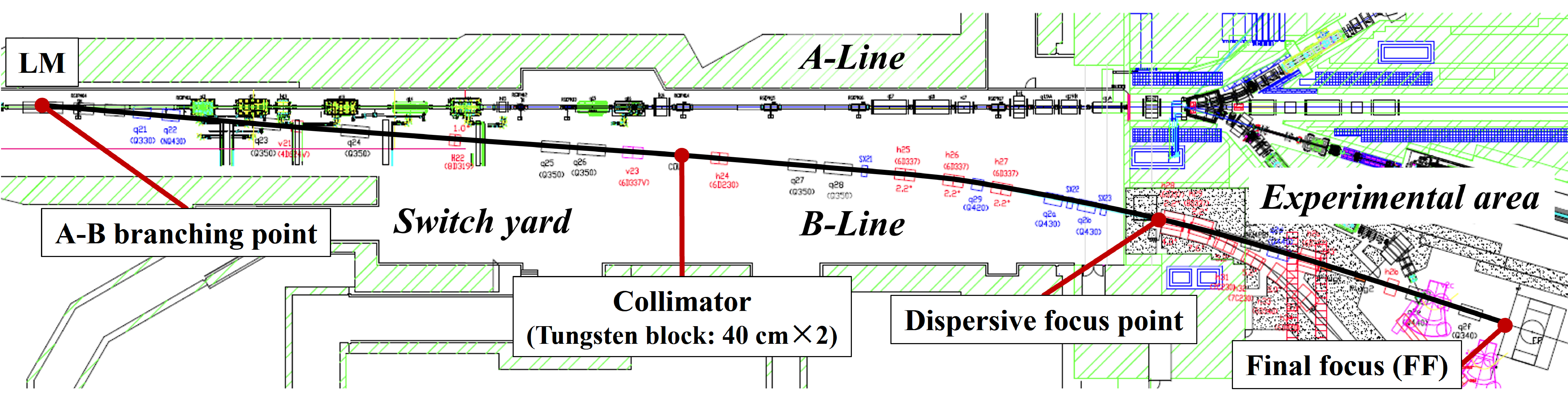}
\caption{
 Schematic view of the B-line under the $\pi20$ mode. 
 The Lambertson magnet (LM) is located at the A-B branching point for secondary beam generation. 
 The downstream focal points are established at the collimator, 
 the dispersive focus, and the experimental target position (FF).
}
\label{fig_pi20}
\end{figure}

Figure~\ref{fig_pi20} shows a schematic view of the B-line. 
We plan to conduct spectroscopy of charmed baryons ($Y_c^{*+}$) via the $\pi^- p \to D^{*-} Y_c^{*+}$ reaction at the B-line~\cite{e50}. 
For this purpose, the optics of the B-line have been designed to deliver secondary beams generated 
at the A-B branching point within a momentum range of $2\text{--}20~{\rm GeV}/c$, 
an operational mode designated as $\pi20$~\cite{HDEFex}. 
The J-PARC Hadron Experimental Facility delivers a slow-extracted proton beam, 
providing a continuous secondary beam during the spill duration rather than a pulsed time structure, 
with a spill duration of $2~{\rm s}$ and a repetition cycle of $4.24~{\rm s}$. 
The beam momentum can be determined from the beam position measurement at the dispersive focus point 
because the beam momentum spread and its horizontal positions are correlated. 
A high momentum resolution of $\Delta p/p \sim 0.1\%~(\sigma)$ is obtained under the $\pi 20$ optics. 
Currently, a beam-splitting septum magnet (Lambertson magnet: LM) is installed at the A-B branching point 
instead of a production target to transfer a small fraction of the primary $30\text{-}{\rm GeV}$ proton beam to the experimental area. 
As a beam loss of several hundred watts is anticipated during the operation of this septum magnet, 
secondary particles are inherently produced at this location. 
Most of the generated secondary particles are pions, with an expected intensity on the order of $10^5~{\rm particles/spill}$. 
Under the current beamline configuration, 
only a positively charged beam can be delivered because the polarities of the beamline magnets 
are fixed to accommodate the primary proton beam.

\begin{figure}[t]
\centering
\includegraphics[width=15cm,clip]{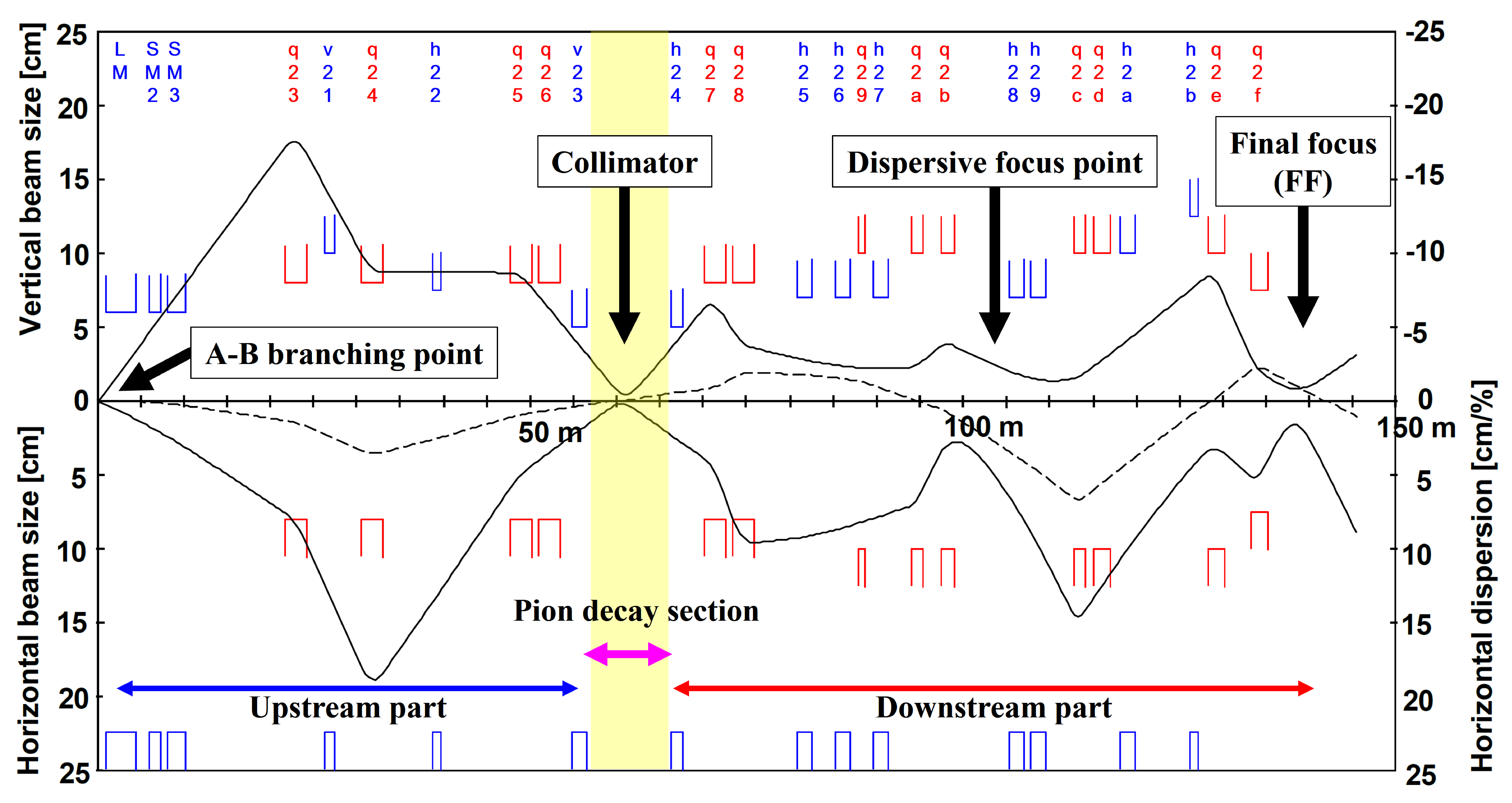}
\caption{
 Calculated beam envelopes (solid lines) and the horizontal dispersion function (dashed line) under the $\pi20$ mode. 
 Blue and red rectangles represent bending and quadrupole magnets, respectively; 
 rectangle dimensions reflect the physical length and aperture (pole gap/width or bore radius) of each element. 
 Upward-emitted pions are captured by the upstream quadrupoles, 
 with their off-axis trajectory compensated by two vertical steering magnets (v21 and v23). 
 The pion decay section for tertiary muon production is labeled, 
 and the generated muons are separated from beam pions via the horizontal bending magnet h24.
}
\label{fig_pi20env}
\end{figure}

The optical design under the $\pi20$ mode was performed 
based on the existing beamline layout for the primary proton beam. 
The calculated beam envelope is shown in Fig.~\ref{fig_pi20env}. 
Since the pole gap of the LM opens upward, 
secondary particles produced at the LM are emitted with an upward vertical angle, 
resulting in an off-axis secondary beam. 
To compensate for this initial trajectory, the beam is first steered downward by the upstream quadrupole magnets 
and subsequently aligned with the central orbit by tuning two downstream vertical steering magnets (v21 and v23). 
The secondary beam is focused 
in both the horizontal and vertical planes at a collimator located approximately $60~{\rm m}$ downstream of the branching point. 
It is horizontally focused again to form a dispersive focus at a position $\sim 105~{\rm m}$ downstream from the branching point. 
Finally, the beam is focused onto the final focus (FF) experimental target situated a further $138~{\rm m}$ downstream.

\subsection{$\mu20$ beamline mode} \label{subsec:mu20_beamline}

\begin{figure}[t]
\centering
\includegraphics[width=15cm,clip]{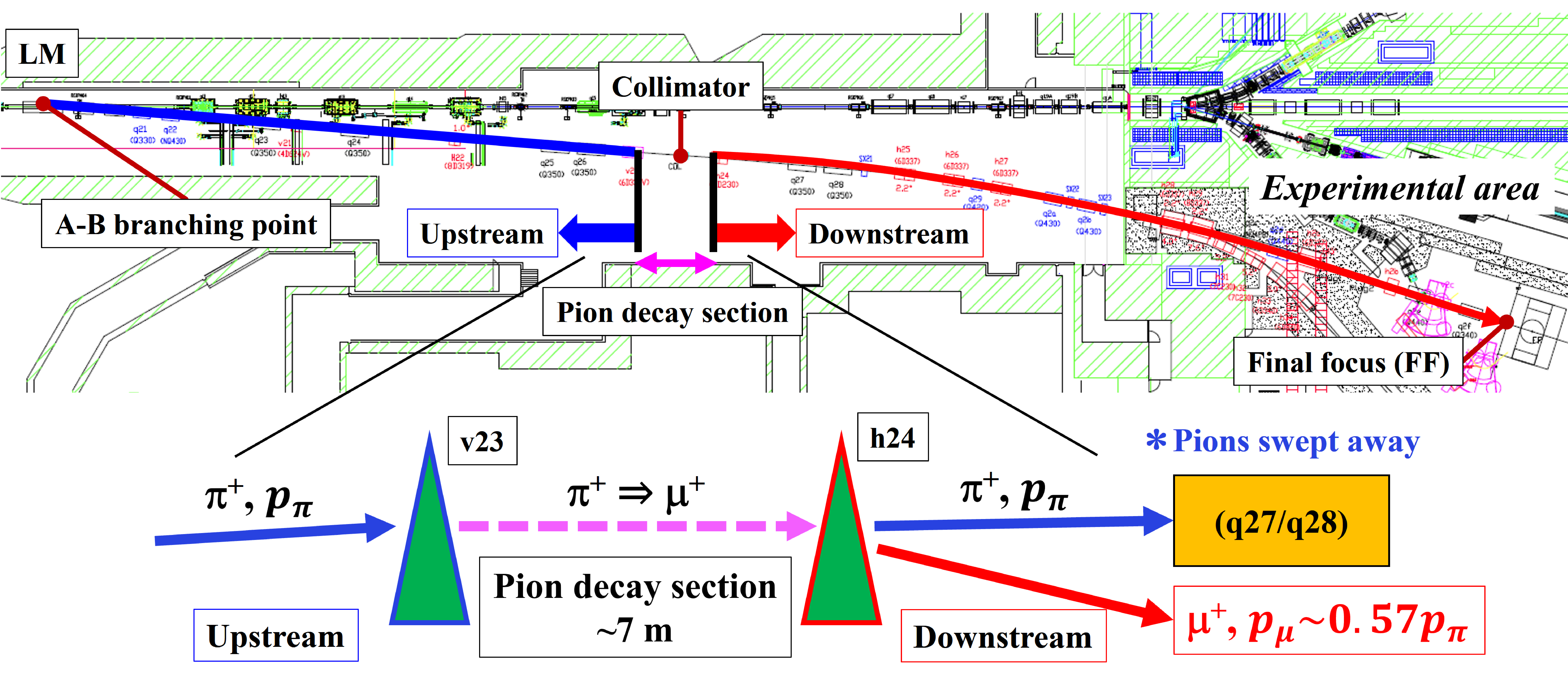}
\caption{
 Concept of the $\mu20$ beam mode. 
 Pions passing through the upstream section of the beamline decay into tertiary muons within the drift space 
 between the vertical bending magnet v23 and the horizontal bending magnet h24. 
 Because the downstream beamline section is operated at $60\%$ of the upstream central momentum, 
 only the backward-decaying muons fall within the acceptance and are transported through the downstream section. 
 In contrast, un-decayed background beam pions are swept away 
 and subsequently absorbed by the downstream magnets in the switchyard.
 }
\label{fig_mu20}
\end{figure}

A pion decays into a muon and a muon neutrino with a mean decay length ($c\tau$) of approximately $7.8~{\rm m}$ in its rest frame. 
A muon emitted backward in the rest frame of a beam pion is boosted forward into the laboratory frame. 
The backward-decaying muon has a momentum of $p_\mu = (m_\mu^2 / m_\pi^2) p_\pi \sim 0.57 p_\pi$ in the relativistic limit, 
a relation practically applicable even at $p_\pi \sim 1~{\rm GeV}/c$. 
Owing to this kinematic property of the decay, 
we have developed an optimized design under the $\mu20$ mode 
to separate the backward-decaying muons from the un-decayed beam pions, as illustrated in Fig.~\ref{fig_mu20}. 
The secondary beam is focused at the collimator located approximately $60~{\rm m}$ downstream of the branching point. 
Vertical and horizontal bending magnets (v23 and h24) are positioned upstream and downstream of the collimator, respectively. 
The maximum opening gaps of the collimator are as wide as $120~{\rm mm}$ in both the horizontal and vertical directions. 
In the $\mu20$ mode, the upstream section of the beamline up to v23 is operated at a specific central momentum, 
while all the magnets (both bending and quadrupole magnets) in the downstream section starting from h24 
are proportionally scaled to $60\%$ of that upstream central momentum to maintain optical alignment. 
The choice of this 60\% scaling factor is based on two-body decay kinematics, 
where a muon emitted in the exact backward direction carries approximately 57\% of the parent pion's momentum. 
Downscaling the downstream beamline section to 60\% thus selectively focuses and transports these backward-decaying muons 
while sweeping away the primary pion background. 
Depending on the target application, 
this scaling factor can be further adjusted to achieve the appropriate balance between muon beam purity and intensity.
Consequently, the backward-decaying muons originating from pion decays within the $\sim 7~{\rm m}$ drift space between v23 and h24 
are efficiently collected and transported through the downstream section. 
In contrast, the un-decayed beam pions are swept away because their trajectories fall outside the acceptance of the downstream section, 
leading to their subsequent absorption by the downstream magnets in the switchyard. 
Under this operating configuration, beam transport simulations verified that the optical profiles 
are well-maintained throughout the scaled sections, 
predicting the extraction of a high-purity muon beam with a momentum bite of $\pm 5\%$~\cite{t106}.

While such a high-momentum muon beam has historically been utilized in major hadron structure experiments, 
such as the COMPASS~\cite{COMPASS} and the upcoming AMBER~\cite{AMBER} programs at CERN, 
those setups traditionally rely on extremely long beamlines to decay secondary pions or kaons and filter out beam impurities. 
In contrast, our approach achieves exceptional beam purity through a distinctly different mechanism: 
simply adjusting the magnet settings optimized for selective muon transport. 
This $\mu20$ mode provides unique opportunities for radiographic imaging with a muon beam in a high-momentum range. 
In the T106 test experiment to be described below,
we measured the intensity, profile and purity of the extracted beam to evaluate its feasibility for practical application to muon radiography.

\section{T106 experiment} \label{sec:t106_experiment}

\subsection{Experimental setup} \label{subsec:experimental_setup}

\begin{figure}[t]
\centering
\includegraphics[width=16cm,clip]{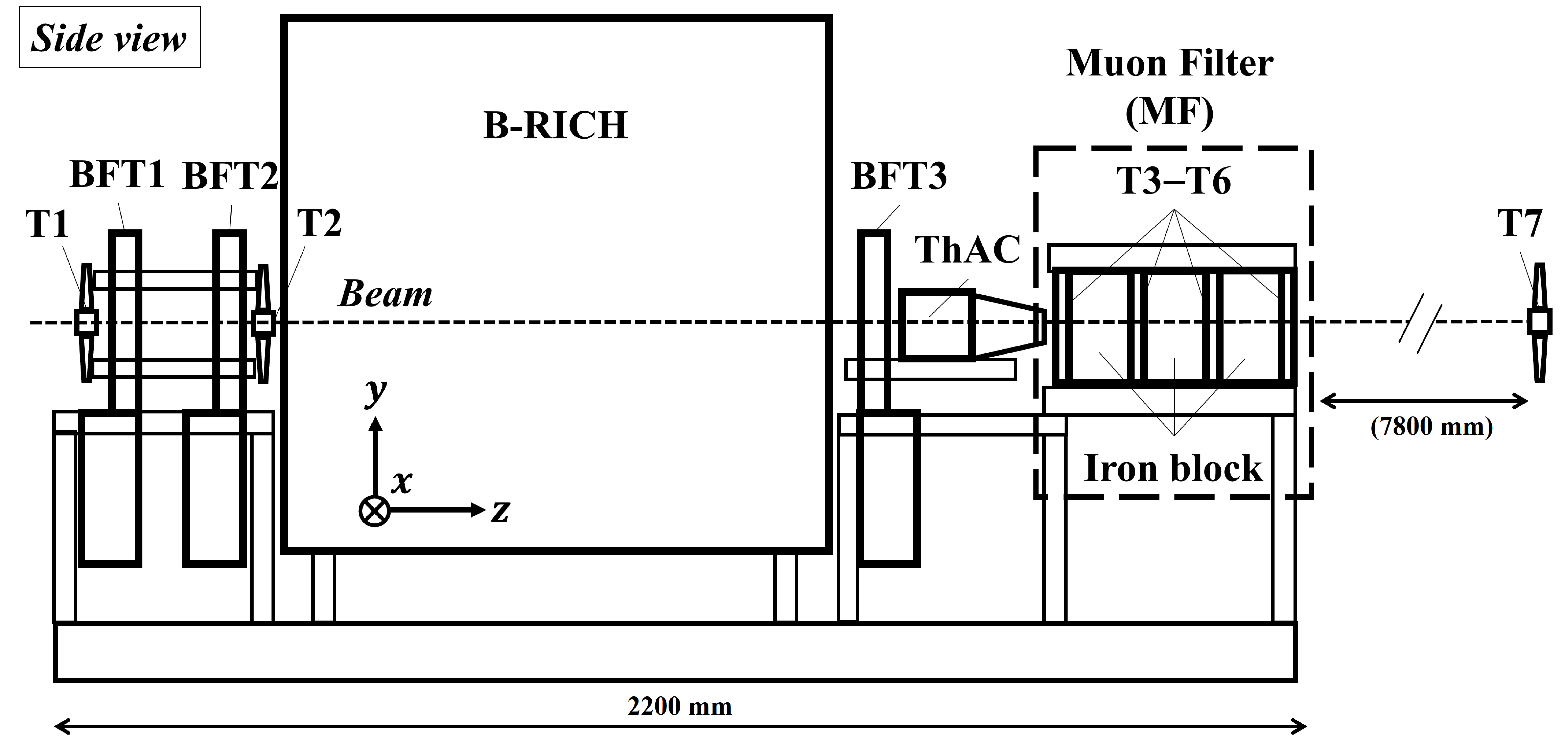}
\caption{
 Schematic view of the experimental detector setup. 
 The coordinate system is defined as a right-handed system where the $z$-axis is oriented along the beam direction, 
 and the $x$- and $y$-axes represent the horizontal and vertical directions, respectively.
}
\label{fig_setup}
\end{figure}

The detector setup was installed downstream of the final quadrupole magnet, q2f, of the B-line. 
The schematic layout of the system is shown in Fig.~\ref{fig_setup}. 
All detector components were aligned along the beam axis ($z$-axis), 
where the $x$- and $y$-axes represent the horizontal and vertical directions, respectively. 
Hodoscope counters (T1--T7) were utilized for timing measurements of beam particles. 
Counters T1, T2, and T7 each comprised a single plastic scintillator (Saint-Gobain Crystals BC-408) 
with an active area of $200 \times 200~{\rm mm}^2$ and a thickness of $10~{\rm mm}$. 
These counters featured metal-package photomultiplier tubes (PMTs), Hamamatsu Photonics H11934, 
coupled to both ends of each scintillator bar. 
Among them, T1 and T2 were primarily used for monitoring the counting rate of beam particles, 
while T7 was positioned downstream to measure the time-of-flight over a long flight path of approximately $7.8~{\rm m}$. 
In contrast, T3--T6 consisted of single plastic scintillators 
with the same geometry ($200 \times 200~{\rm mm}^2$ active area and $10~{\rm mm}$ thickness), 
but were read out by a 2-inch PMT (Hamamatsu Photonics H6140) connected to a single end. 
These active areas were chosen to ensure full geometric acceptance 
for the expected beam envelope derived from the transport simulations. 
A muon filter (MF) system was configured using T3--T6 and three iron blocks arranged alternately. 
To match the beam profile coverage, 
the iron blocks featured an active area of $200 \times 200~{\rm mm}^2$ and a thickness of $100~{\rm mm}$, 
yielding a total iron thickness of $300~{\rm mm}$ to serve as a particle identification system.

For the tracking system, three beam fiber trackers, designated as BFT1, BFT2, and BFT3, were employed. 
Each BFT was constructed as a three-layer scintillating-fiber tracker covering an effective area of $140 \times 200~{\rm mm}^2$, 
utilizing scintillating fibers (Kuraray SCSF-78M) arranged in a staggered configuration to construct a single layer. 
The fiber diameters for BFT1 and BFT2 were $0.5~{\rm mm}$, whereas BFT3 utilized $1.0~{\rm mm}$ fibers. 
Within each tracker, the fiber layers were oriented at tilt angles of $0^\circ$ ($x$), $+30^\circ$ ($u$), and $-30^\circ$ ($v$) 
with respect to the vertical direction ($y$-axis). 
Each individual fiber was optically coupled via air 
contact to a corresponding channel of an $8 \times 8$ multi-pixel photon counter (MPPC) array (Hamamatsu Photonics S13360-1350) 
with a channel size of $1.3 \times 1.3~{\rm mm}^2$. 
BFT1 and BFT2 were primarily used for the beam profile measurements.

Beam particles were identified using a beam ring-imaging Cherenkov detector (B-RICH)~\cite{brich} 
and a threshold-type aerogel Cherenkov detector (ThAC)~\cite{t106proc2}. 
These particle identification (PID) detectors were dedicated 
exclusively to the identification of beam pions ($\pi$), kaons ($K$), and protons ($p$) during the hadron-beam evaluation measurements. 
The B-RICH detector utilized silica aerogel with a refractive index of $1.02$. 
The generated Cherenkov light was reflected by a spherical mirror 
and focused onto an MPPC array (Hamamatsu Photonics S13361-3050AE-04). 
This B-RICH system provides sufficient performance to discriminate between $\pi$ and $K$ 
in the beam momentum range from $5.0$ to $8.5~{\rm GeV}/c$, 
as well as $\pi$ and $p$ at a beam momentum of $10~{\rm GeV}/c$. 
The ThAC detector utilizing silica aerogel with a refractive index of $1.007$ was employed 
to identify these particle species by distinguishing whether or not the particle velocity exceeded the Cherenkov emission threshold. 
Specifically, the Cherenkov momentum thresholds at this refractive index 
are calculated to be $1.18$, $4.17$, and $7.92~{\rm GeV}/c$ for $\pi$, $K$, and $p$, respectively. 
In the ThAC configuration, a light concentrator was positioned downstream of the aerogel radiator 
to efficiently collect the emitted Cherenkov light onto the photon readout system, 
which utilized an MPPC array (Hamamatsu Photonics S13361-3075AE-08).

\subsection{Data acquisition system} \label{subsec:daq_system}

Data acquisition was handled by a trigger-less streaming-readout data acquisition (SRODAQ) system developed 
and established through our recent research~\cite{t103, SRODAQ}. 
The SRODAQ architecture continuously acquired the entire data stream 
from time-synchronized front-end electronics~\cite{mikumari} directly into a computing node. 
To construct this network-based system, 
AMANEQ modules equipped with streaming high-resolution TDCs (HR-TDC)~\cite{hr-tdc} 
were utilized for the timing hodoscopes and the ThAC detector, 
while CIRASAME modules~\cite{t103,cirasame}, designed for reading silicon photomultipliers (SiPM) 
and incorporating streaming low-resolution TDCs, were employed for the BFT and B-RICH readouts. 
The data stream was managed using the NestDAQ software framework~\cite{nestdaq}. 
Although the beam intensity during this test experiment was on the order of $10^5$ particles per beam spill, 
we utilized both the trigger-emulation mode implemented in the time synchronization module 
and the online data filtering logic within the NestDAQ software processes. 
Under both processing schemes, the $T1 \times T2$ coincidence was shared as a common criterion; 
it was applied as a hardware logic signal to the module for the trigger-emulation mode, 
whereas it was executed as a software process for the online data filtering logic. 
This filtering configuration provided a sufficiently unbiased sample, 
thereby enabling a successful evaluation of both the muon beam data and the beam background. 
With this operational setup, the system successfully recorded all the beam particle data without introducing dead-time losses.

\subsection{Beam measurements and conditions} \label{subsec:beam_measurements_and_conditions}

The T106 experiment characterized the beam properties for both hadrons and muons. 
We investigated beam momentum settings of $3$, $5$, and $10~{\rm GeV}/c$ under the $\pi20$ mode. 
At the beginning of the experiment, a beamline magnet parameter scan was performed under the $\pi20$ mode 
to center the beam position and maximize its intensity. 
After completing the beamline tuning with the hadron beam, 
the magnet configurations were switched to the $\mu20$ mode to evaluate the properties of the delivered muon beams. 
Under the $\mu20$ mode, the beam intensity, profile, and purity were systematically evaluated 
by scaling the momentum setting of the downstream section to $60\%$, $75\%$, $80\%$, and $85\%$ 
of the upstream central momenta ($3$, $5$, and $10~{\rm GeV}/c$). 
This specific scanning range was strategically selected based on the kinematics of two-body pion decay. 
Scaling factors below $60\%$ were not explored because they fall significantly below the kinematic peak of the backward-decaying muons ($\sim 57\%$), 
where the expected muon intensity drops drastically and low-energy hadron backgrounds dominate. 
Conversely,  $85\%$ was chosen as the upper limit 
because this value corresponds to the laboratory momentum of muons emitted at $90^\circ$  in the center-of-mass frame. 
In total, twelve distinct operating configurations were successfully measured during the T106 experiment, 
establishing a comprehensive dataset for the performance verification of the beamline.

\section{Results} \label{resultsT106}

The intensity, profile, and purity of the delivered muon beams were systematically evaluated based on the recorded datasets. 
The beam intensity was determined using hardware scalers measuring the count rate of the $T1 \times T2$ coincidence. 
To evaluate the beam profile, linear tracking via BFT1 and BFT2 
was utilized to reconstruct the horizontal ($x$) and vertical ($y$) positions alongside the incident angle distributions. 
The spatial beam profile at the T3 hodoscope position was subsequently obtained 
by extrapolating these reconstructed trajectories to a position $1.40~{\rm m}$ downstream from the first layer of BFT1.

The systematic evaluation of beam purity was performed utilizing the MF system, 
where the iron blocks featuring an active area of $200 \times 200~{\rm mm}^2$ and a total thickness of $300~{\rm mm}$ 
were arranged alternately with the T3--T6 hodoscope counters.
This alternating structure serves as a particle identification system to separate tertiary muons from hadron background events. 
While beam pions and other background hadrons are heavily attenuated 
or absorbed through hadronic interactions within the thick iron blocks, 
high-energy muons penetrate the entire MF assembly as minimum ionizing particles. 
Consequently, by analyzing the signal attenuation characteristics across T3 to T6, 
the MF system achieves a clear identification of the extracted muon beam.

\subsection{Intensity} \label{results:intensity}

\begin{table}[t] 
\centering
\caption{
 Summary of the obtained hadron beam intensity and particle abundances. 
 The listed beam intensities represent the average values measured over ten beam spills.
}
\begin{tabular}{cccc}
\hline
Momentum~[${\rm GeV}/c$] & 3 & 5 & 10 \\
\hline
Intensity~[M/spill] & 0.2 & 0.6 & 1.4 \\
Abundance~($\pi \colon p$) & $0.66 \colon 0.34$ & $0.60 \colon 0.40$ & $0.39 \colon 0.61$ \\
\hline
\end{tabular}
\label{tab_intensity_had}
\end{table}

Table~\ref{tab_intensity_had} shows the hadron beam intensity obtained 
after the beamline magnet parameter scan under the $\pi20$ mode. 
The hadron beam comprised various generated particles, including pions, kaons, and protons. 
Due to their production cross-sections, pions and protons were the dominant components; 
thus, their respective beam abundances are summarized in Table~\ref{tab_intensity_had}. 
The beam pions and protons at central momenta of $3$ and $5~{\rm GeV}/c$ were identified by the ThAC, 
whereas the B-RICH was utilized for particle identification at $10~{\rm GeV}/c$~\cite{t112}. 
Under the $\pi20$ mode, a typical beam intensity on the order of $10^5$ particles per spill 
was delivered (with a spill duration of $2~{\rm s}$ and a repetition cycle of $4.24~{\rm s}$). 
Table~\ref{tab_intensity_muon} summarizes the muon beam intensity measured after switching to the $\mu20$ mode, 
including the intensities across all downstream scaled modes. 
Although the forward Lorentz boost inherently limits the laboratory-frame acceptance for backward-decaying muons, 
a small fraction of residual hadron contamination was still anticipated. 
Under the $\mu20$ mode, the resulting muon beam intensity achieved was on the order of $10^3$ particles per spill.

\begin{table}[t]\centering
\caption{
 Summary of the obtained muon beam intensity in units of ${\rm k/spill}$. 
 The central momentum settings for both the upstream and downstream scaled sections are listed. 
 The beam intensities represent the average values measured over ten beam spills.
}
\begin{tabular}{ccccc}
\hline
$p_{\rm{upstream}}$ & \multicolumn{4}{c}{Downstream scaling factor} \\
$[{\rm GeV}/c]$ & 60\% & 75\% & 80\% & 85\% \\
\hline
3 & 0.42 & 0.64 & 1.0 & 2.0 \\
5 & 1.7 & 2.6 & 3.7 & 8.3 \\
10 & 2.4 & 4.6 & 6.6 & 17.0 \\
\hline
\end{tabular}
\label{tab_intensity_muon}
\end{table}

\subsection{Profile} \label{results:profiles}

\begin{figure}[t]
\centering
\includegraphics[width=16.0cm,clip]{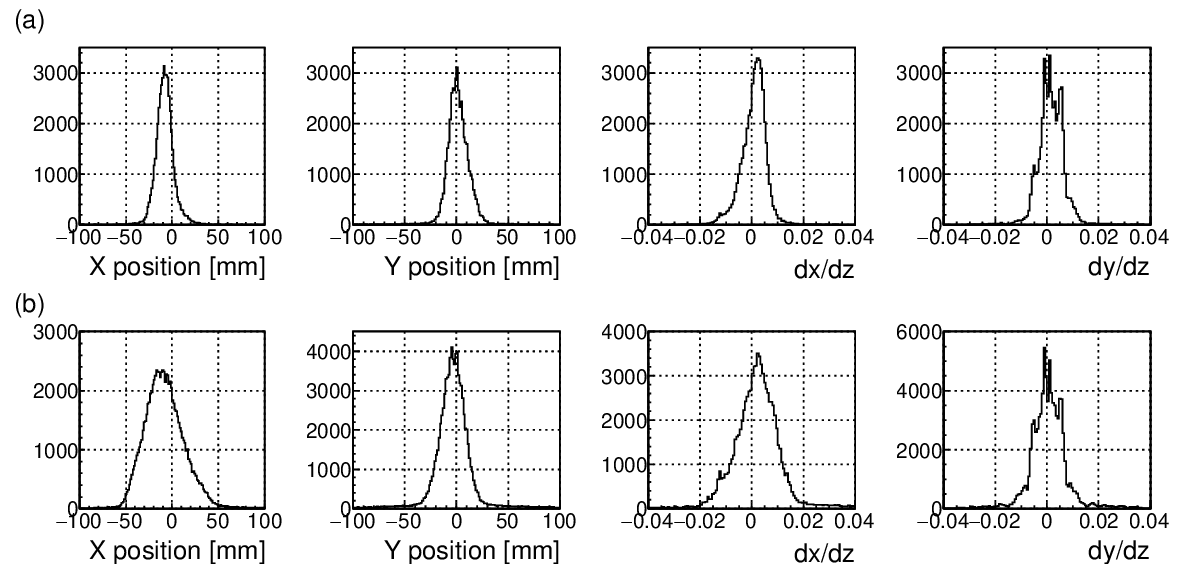}
\caption{
 Spatial profiles and incident angle distributions of the beams evaluated at the T3 position: 
 (a) horizontal ($x$) and vertical ($y$) spatial profiles of the hadron beam 
 along with their respective incident angle distributions ($dx/dz$, $dy/dz$) 
 obtained at a central beam momentum of $5~{\rm GeV}/c$ under the $\pi20$ mode, 
 and (b) corresponding profiles of the tertiary muon beam obtained 
 with a downstream scaling factor of $60\%$ (delivering $3~{\rm GeV}/c$) under the $\mu20$ mode.
}
\label{fig_profile}
\end{figure}

Figure~\ref{fig_profile}(a) shows the horizontal ($x$) and vertical ($y$) spatial profiles of the hadron beam at the T3 position, 
along with their respective incident angle distributions ($dx/dz$, $dy/dz$), 
obtained at a central beam momentum of $5~{\rm GeV}/c$ under the $\pi20$ mode. 
At this momentum, the spatial and incident angle distributions relative to the beamline axis 
were found to be approximately $10~{\rm mm}$ and $4~{\rm mrad}$ in standard deviation ($\sigma$), respectively. 
Similar spatial beam sizes were obtained for the $3$ and $10~{\rm GeV}/c$ hadron beam settings. 
The characteristic structures observed in the vertical incident angle distribution ($dy/dz$) 
reflect the geometrically asymmetric footprint of the secondary beam transport from the LM. 
A slight deviation of the hadron beam profile center from the nominal beamline axis was observed at the T3 position; 
however, this discrepancy is attributed to a minor operational alignment 
that can be straightforwardly optimized by adjusting the final steering magnets, 
and it does not affect the intrinsic beam quality or the core conclusions of this study. 
Therefore, the beam configuration was maintained without further local tuning during the present evaluation. 
Figure~\ref{fig_profile}(b) shows the spatial profiles of the muon beam obtained at the T3 position 
with a downstream scaling factor of $60\%$ under the $\mu20$ mode. 
These profiles were inherently broader than those of the hadron beam 
due to the kinematic characteristics associated with collecting the backward-decaying muons; 
both the spatial and angular $\sigma$ widths relative to the beamline axis 
were approximately $1.5$ times larger than those of the hadron beam.
Crucially, across all twelve investigated operating modes, 
encompassing the three upstream central momenta and the four downstream scaling factors, 
the spatial and angular beam profile characteristics consistently exhibited the same systematic tendencies.

\subsection{Purity} \label{results:purity}

The muon beam purity under the $\mu20$ mode was quantitatively evaluated 
by analyzing the beam attenuation characteristics across the MF layers. 
Ideally, a pure muon beam at the present energy scale completely penetrates the MF without significant attenuation. 
In contrast, a background hadron beam undergoes sequential attenuation via nuclear interactions 
as it propagates into the deeper layers of the MF. 
By normalizing the beam tracking counts at the entrance layer (T3) to unity, 
the muon beam purity was extracted from the attenuation gradients observed from T4 to T6. 
For the beam analysis, the event selection required the horizontal and vertical track positions on the T3 plane 
to be within $|x| < 50~{\rm mm}$ and $|y| < 50~{\rm mm}$, respectively. 
This spatial constraint was applied to ensure that the beam profile was sufficiently smaller 
than the physical acceptance of the T3--T6 layers and the interleaved iron blocks, 
thereby effectively suppressing beam leakage from the side walls of the MF. 
Prior to the purity evaluation, the intrinsic detection efficiencies of the MF timing layers (T3--T6) 
were experimentally determined using the data obtained in the $60\%$ downstream scaled mode 
at the upstream beam momentum of $5~{\rm GeV}/c$. 
By requiring coincidence hits in the upstream triggers ($T1 \times T2$) and the hindmost downstream counter (T7), 
muon beam events completely penetrating the MF could be cleanly isolated. 
Under this selection, the intrinsic efficiencies for the individual counters from T3 to T6 
were determined to be $99.84 \pm 0.02\%$, $98.51 \pm 0.05\%$, $99.54 \pm 0.03\%$, and $99.67 \pm 0.02\%$, respectively.
Although the efficiency of T4 was found to be approximately $1\%$ lower than that of the other layers, 
potentially due to a minor optical coupling variance or an electrical contact issue between the plastic scintillator and the photomultiplier tube, 
this variation was explicitly corrected for by scaling the raw particle counts with the measured efficiencies in the subsequent analysis stages. 
In this analysis, minimum ionizing particle (MIP) selection criteria were established for T3--T6 
using the data obtained from the $60\%$ downstream scaled mode, 
where an energy loss of at least half of the MIP peak value was required to define a particle hit. 
This established MIP selection was subsequently applied across all following analyses to ensure a consistent counting of hits in T3--T6.

\begin{figure}[t]
\centering
\includegraphics[width=16.0cm,clip]{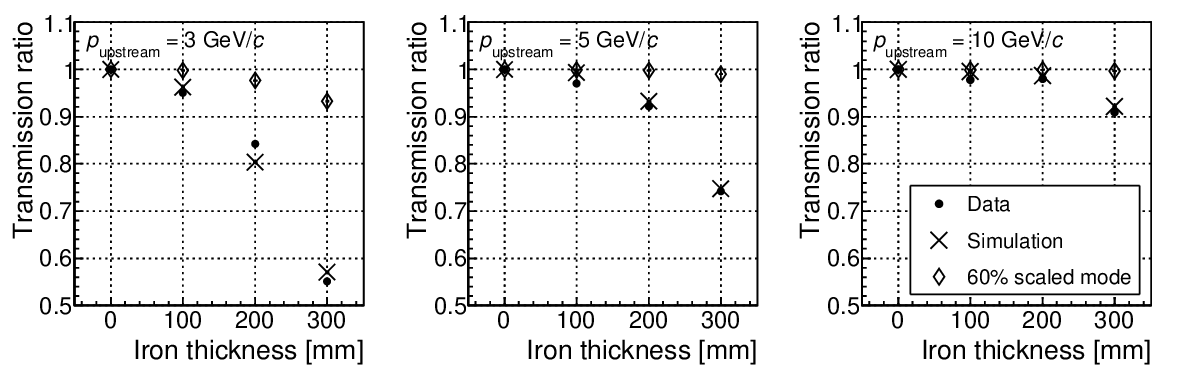}
\caption{
 Beam attenuation profiles measured at upstream central beam momenta of $3$, $5$, and $10~{\rm GeV}/c$ 
 as a function of the total iron thickness, normalized to the T3 counts. 
 The vertical axis represents the beam transmission ratio, 
 scaled from $0.5$ to $1.1$ to clearly encompass all data points. 
 The experimental data (solid circles, crosses, and open diamonds) are superposed 
 against the corresponding {\sc Geant4} simulation results as indicated in the legend, 
 highlighting the suppressed attenuation in the $60\%$ downstream scaled mode.
}
\label{fig_attenuation_hadron}
\end{figure}

Next, the detector response to a pure hadron beam was investigated. 
The beam attenuation profiles across the MF layers measured at initial central beam momenta of $3$, $5$, and $10~{\rm GeV}/c$ 
under the $\pi20$ mode are shown in Fig.~\ref{fig_attenuation_hadron}, normalized to the T3 counts. 
For lower beam momenta, a clear trend of enhanced attenuation in the transmission ratio 
was observed as the particles progressed into the deeper layers. 
Figure~\ref{fig_attenuation_hadron} 
also superposes the experimental data obtained in the $60\%$ scaled mode, 
demonstrating that the beam attenuation under the $\mu20$ mode is substantially suppressed compared to that of the raw hadron beam. 
The Monte Carlo simulations were conducted using the {\sc Geant4} toolkit (version 10.7)~\cite{geant4}, 
wherein the exact geometric configuration of the MF was replicated. 
The geometry modeled four timing detectors ($200 \times 200 \times 10~{\rm mm}^3$) interleaved with three iron absorber segments, 
each composed of four cast iron blocks (twelve blocks in total) to form a unified absorber structure 
with a total thickness of $300~{\rm mm}$. 
Because the utilized absorber blocks consist of cast iron, they possess a characteristically low material density. 
Based on direct measurements of the mass and dimensions of the actual cast iron pieces, 
their densities exhibited structural variations ranging from $6.8$ to $8.0~{\rm g/cm^3}$; 
however, a global average value of $7.23~{\rm g/cm^3}$ was adopted in the simulation geometry. 
Notably, any potential systematic uncertainties introduced by this local density variation do not affect the integrity of our conclusions, 
as the simulation framework was rigorously validated through a direct comparison with the empirical hadron beam data. 
Specifically, for the analysis of the $60\%$ downstream scaled mode at an upstream central momentum of $5~{\rm GeV}/c$, 
the delivered backward-decaying muons possess a central momentum of $3~{\rm GeV}/c$. 
Consequently, the pure hadron beam data experimentally measured at $3~{\rm GeV}/c$ 
can be directly utilized as a reliable empirical template to cross-check and validate the detector response and attenuation profiles, 
thereby minimizing the reliance on the absolute density precision in the simulation framework.

For the beam optical input, the initial spatial and angular distributions were generated as independent Gaussian profiles. 
The spatial parameters for each simulation were derived from the specific $\sigma$ values 
experimentally measured at each of the twelve distinct operating configurations.
Because the actual beam profiles featured a sufficiently small spatial size and a narrow angular divergence 
relative to the physical dimensions of the hodoscope counters and the iron blocks, 
this independent profile approximation was thoroughly adequate 
without requiring a rigorous replication of exact phase-space correlations. 
The simulated physics processes incorporated particle decay, electromagnetic interactions, and comprehensive hadronic processes. 
For hadronic elastic modeling, the native {\sc Geant4} hadron elastic scattering model was utilized; 
specifically, the $\pi^{\pm}$ components were modeled via high-energy hadron-nucleus elastic scattering (\texttt{HadrNucleusHE}) 
at energies above $1~{\rm GeV}$, 
whereas the proton component utilized the Chiral Invariant Phase Space (\texttt{CHIPS}) elastic scattering model. 
For inelastic processes, string and cascade models were selectively applied depending on the energy regime: 
the Fritiof string model (\texttt{FTF}) was adopted for the higher-energy region ($> 4~{\rm GeV}/c$), 
and the Bertini cascade model (\texttt{BERT}) was employed for the lower-energy region ($< 5~{\rm GeV}/c$). 
As a result, the {\sc Geant4} simulation successfully reproduced the experimental attenuation profiles of the hadron beam.
The initial beam composition ratios injected into the simulation 
were constrained by the empirical data presented in Sec.~\ref{results:intensity}.

The extraction of the muon beam purity was performed using the responses of the muon and hadron components 
derived from both the experimental data and the simulations. 
Because high-energy muons completely penetrate the filter with negligible attenuation, 
any observed attenuation deviation in the experimental data is uniquely driven by the accompanying hadron contamination. 
The actual beam is modeled as a linear combination of these two components. 
The true muon purity $R$ was determined by minimizing the residual sum of squares ($\mathrm{RSS}$) 
between the experimental hodoscope data and the simulated response across the MF layers:
\begin{equation}
\mathrm{RSS} = \sum_{i} \left( \mathrm{Data}_i - \left[ R \cdot \mathrm{Sim}_{i}^{\mu} + (1-R) \cdot \mathrm{Sim}_{i}^{\mathrm{had}} \right] \right)^2,
\end{equation}
where $i$ denotes the index of the MF timing layers (T3--T6), 
and $\mathrm{Data}_i$, $\mathrm{Sim}_{i}^{\mu}$, and $\mathrm{Sim}_{i}^{\mathrm{had}}$ 
represent the normalized counts at the $i$-th layer for the experimental data, 
the simulated muon response, and the simulated hadron response, respectively. 
The hadronic response $\mathrm{Sim}_{i}^{\mathrm{had}}$ itself is constructed as a composite of pion and proton responses. 
The specific hadron contamination ratios ($\pi^+ : p$) utilized in the {\sc Geant4} simulation 
were directly constrained by the experimental beam abundances measured under the $\pi20$ mode. 
The $\mathrm{RSS}$ minimization was adopted for evaluating the muon purity 
because it effectively mitigates localized detector systematic fluctuations through a global attenuation fit, 
mathematically converges with maximum likelihood estimation under high-statistics conditions, 
and ensures a stable, unique global optimum for the linear combination model.
Furthermore, for the $5~{\rm GeV}/c$ upstream configuration at $60\%$ downstream scaling, 
the $3~{\rm GeV}/c$ hadron beam data was available to serve as a direct experimental template. 
This allowed us to cross-evaluate the muon purity by substituting the simulated hadron response 
with the purely empirical hadron response.

\begin{figure}[t]
\centering
\includegraphics[width=16.0cm,clip]{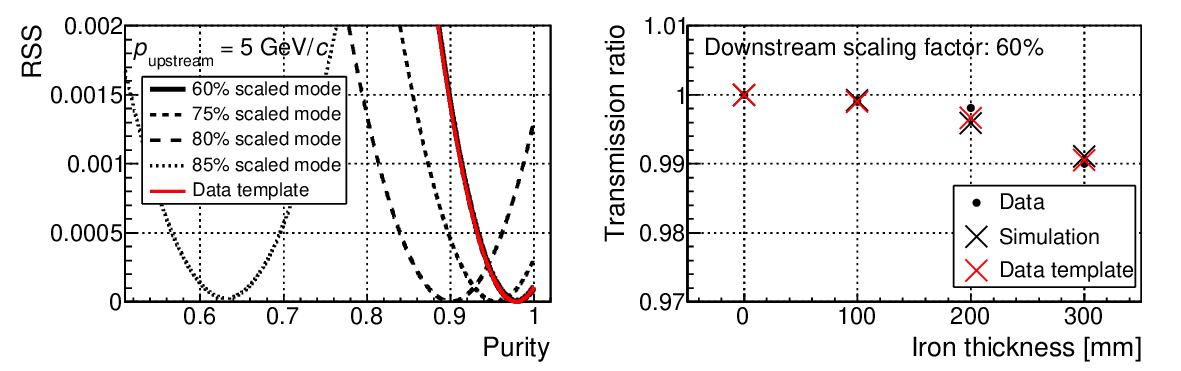}
\caption{
 Extraction results of the muon beam purity for the $5~{\rm GeV}/c$ operating configurations: 
 (a) the $\mathrm{RSS}$ curves obtained using the {\sc Geant4} templates for each downstream scaling factor, 
 alongside the empirical template curve utilizing the $3~{\rm GeV}/c$ hadron data, 
 and (b) the beam transmission ratio for the $60\%$ downstream scaled mode. 
 The specific meaning of each symbol and color is indicated in the legend. 
 In panel (b), the vertical axis is scaled from $0.97$ to $1.01$ to clearly encompass all data points in the proximity of unity.
}
\label{fig_rss}
\end{figure}

\begin{figure}[t]
\centering
\includegraphics[width=16.0cm,clip]{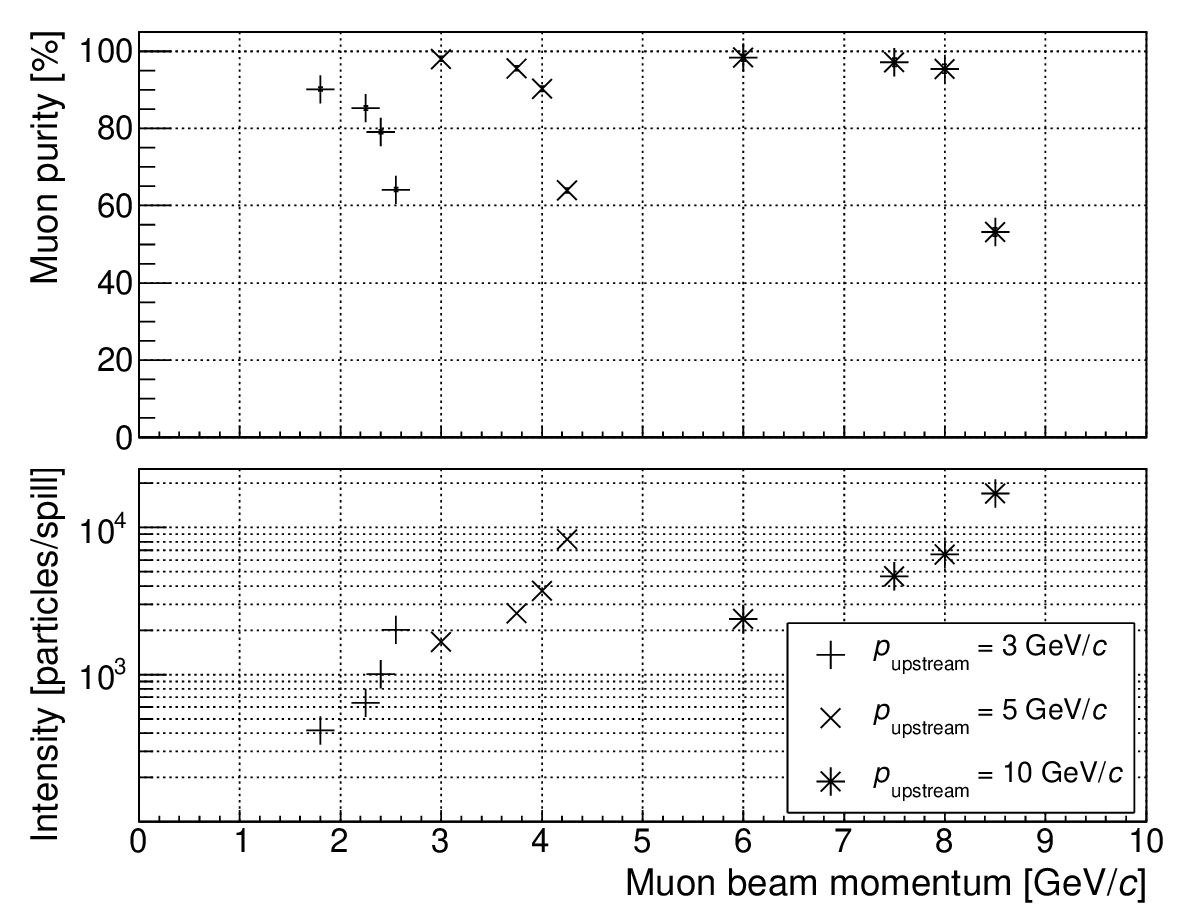}
\caption{
 Summary of the evaluated muon beam intensity and purity as a function of the downstream central momentum setting 
 for the upstream central beam momenta of $3$, $5$, and $10~{\rm GeV}/c$. 
 The plotted data points systematically encompass the twelve distinct operating configurations measured under the $\mu20$ mode.
}
\label{fig_muon_summary}
\end{figure}

Figure~\ref{fig_rss} illustrates the extraction results for the $5~{\rm GeV}/c$ configurations. 
The unique global minimum of the ${\rm RSS}$ curve represents the extracted muon beam purity, 
which was systematically evaluated for the $60\%$, $75\%$, $80\%$, and $85\%$ downstream scaling factors. 
For the $60\%$ scaled mode, the purity extracted using the {\sc Geant4} hadronic template was determined to be $98.0 \pm 0.6\%$, 
whereas the purity evaluated using the empirical $3~{\rm GeV}/c$ hadron data yielded $97.9 \pm 0.9\%$. 
Because the experimental hadron template represents the direct detector response to actual beam particles, 
essentially free from any structural assumptions in the simulation, 
the agreement between these two independently extracted values 
firmly validates the accuracy and reliability of our simulation-based evaluation framework. 
The identical systematic attenuation and template-fitting analysis 
was straightforwardly applied to the other central momentum configurations ($3$ and $10~{\rm GeV}/c$), 
confirming the expected beam characteristics. 
Table~\ref{tab_muon_purity} summarizes the extracted muon purities across all measured downstream scaled modes 
for the initial central beam momenta of $3$, $5$, and $10~{\rm GeV}/c$. 
Figure~\ref{fig_muon_summary} presents the summary of the muon beam intensity and the corresponding purity evaluated 
as a function of the downstream beam momentum under the $\mu20$ mode.

\begin{table}[t]
\centering
\caption{
 Summary of the obtained muon beam purity in units of $\%$. 
 The central momentum settings for both the upstream and downstream scaled sections are listed. 
 The presented purity values correspond to the twelve distinct operating configurations measured under the $\mu20$ mode.
}
\begin{tabular}{ccccc}
\hline
$p_{\rm{upstream}}$ & \multicolumn{4}{c}{Downstream scaling factor} \\
$[{\rm GeV}/c]$ & 60\% & 75\% & 80\% & 85\% \\
\hline
3 & $90.2\pm0.5$ & $85.2\pm0.5$ & $79.1\pm0.5$ & $61.4\pm0.5$ \\
5 & $98.0\pm0.6$ & $95.6\pm0.7$ & $90.3\pm0.6$ & $63.9\pm0.6$ \\
10 & $98.3\pm0.9$ & $97.1\pm1.1$ & $95.4\pm1.1$ & $53.2\pm1.1$ \\
\hline
\end{tabular}
\label{tab_muon_purity}
\end{table}

\section{Discussion}

The experimental evaluations demonstrated the fundamental capabilities of the high-momentum muon beamline at J-PARC. 
However, the resulting muon beam intensity delivered under the $\mu20$ mode was on the order of $10^3~{\rm particles/spill}$. 
While the current intensity is sufficient for proof-of-principle tests and baseline performance validation, 
practical application to high-resolution muon radiography for massive, large-scale structures 
imperatively requires significantly higher beam statistics. 
To meet this statistical requirement, 
a future upgrade of the B-line is planned to install a $15\text{-}{\rm kW}$ loss production target at the branching point, 
which is substantially higher than the current loss of several hundred watts at the LM. 
Consequently, the beam intensity is expected to increase by more than two orders of magnitude 
compared to the current experimental conditions, 
reaching over $10^5~{\rm particles/spill}$ under the $\mu20$ mode. 
This enhancement will extend the availability of the beamline, 
thereby opening up new avenues for various particle physics and interdisciplinary imaging applications.

Regarding the spatial characteristics, the spatial and angular profiles of the tertiary muon beam 
remain sufficiently well-constrained to feasibly perform muon radiography experiments. 
For practical radiography applications, we plan to adjust the excitation parameters of the final quadrupole magnets 
in the beamline to conduct dedicated beam tuning. 
This operational flexibility will allow us to either expand the spatial beam size or minimize the angular divergence 
according to the specific physical dimensions and geometric constraints of the imaging target. 
While a comprehensive experiment for the quantitative evaluation of the imaging performance, 
such as the exact spatial resolution and material penetrability for specific large-scale objects, 
is proposed to be demonstrated as a dedicated radiography experiment~\cite{t113}, 
the robust beam qualities achieved in the present work establish a highly promising and reliable foundation for these future applications.

The quantitative verification of the muon purity firmly establishes the performance and reliability of the $\mu20$ mode. 
Most notably, operating at a backward-decay scaling factor of $60\%$ consistently provided a high-purity muon beam delivery configuration. 
This particular operating scheme successfully achieved a substantial suppression of parent hadron backgrounds 
while maintaining a viable beam fluence, demonstrating the practical efficacy of the $\mu20$ mode. 
The systematic outcomes across various downstream momentum settings 
elucidate the physical trade-off inherent in the kinematic selection: 
tuning toward backward-decay angles (lower scaling factors) suppresses hadronic contamination 
at the expense of the absolute beam intensity, aligning precisely with beam optics expectations. 
Conversely, opting for higher scaling factors enhances the overall intensity while compromising the muon purity. 
The high purity achieved at the $60\%$ scaling setting demonstrates that the beamline framework 
is capable of effectively suppressing the un-decayed primary hadron background. 
These comprehensive results confirm that the established beamline 
successfully delivers a high-momentum muon beam with exceptional purity, 
thereby paving the way for diverse applications that were previously unfeasible.

\section{Conclusion}

We have successfully realized and characterized a novel high-purity muon beam delivery scheme, 
designated as the $\mu20$ mode, at the high-momentum hadron beamline (B-line) within the J-PARC Hadron Experimental Facility. 
Operating in the straight section of the channel, 
this scheme provides tertiary muons originating from the decays of secondary pion beams, 
with the beam optics optimized for backward-decay kinematics to efficiently eliminate parent hadron backgrounds. 
To systematically investigate the operational versatility and performance limits of this framework, 
comprehensive parametric characterizations were executed across twelve distinct operating modes, 
comprising three upstream central beam momenta of $3$, $5$, and $10~{\rm GeV}/c$ 
and four downstream momentum scaling factors of $60\%$, $75\%$, $80\%$, and $85\%$.

Regarding the core performance, 
benchmarks for the configuration with a $5~{\rm GeV}/c$ upstream momentum operating at a $60\%$ downstream scaling factor, 
delivering a $3~{\rm GeV}/c$ muon beam, achieved a stable standalone muon flux on the order of $10^3~{\rm particles/spill}$. 
Under this specific condition, the beam was transported with a highly tailored profile, 
resulting in a spatial $\sigma$ spread of approximately $15~{\rm mm}$ at the downstream detector position.
Additionally, the systematic attenuation methodology utilizing the iron-absorber detector system 
verified an exceptional muon purity of $98.0 \pm 0.6\%$ for this $3~{\rm GeV}/c$ muon beam via the hadron beam template analysis, 
which was further confirmed to be $97.9 \pm 0.9\%$ through independent validation using the empirical hadron data, 
demonstrating excellent mutual agreement. 
Furthermore, the systematic evaluations verified that the other measured momentum configurations 
consistently achieved comparable beam intensities and high muon purities.

These studies successfully establish the $\mu20$ mode within this high-momentum beamline, 
thereby providing a robust platform for diverse advanced applications, including high-quality muon radiography. 
Notably, this comprehensive study demonstrates the first successful operation of a high-momentum and high-purity muon beamline 
within an experimental facility in Japan.

\section*{Acknowledgments}

This work was supported by JST [K program] Japan Grant Number JPMKP24J3, 
Japan Society for the Promotion of Science (JSPS) Grant-in-Aid for Scientific Research (S) Grant No. JP22H04940,
and Grant-in-Aid for Scientific Research (A) Grant No. 22H00124.
The authors thank the members of the J-PARC Hadron Beam Line Group 
for their dedicated efforts to deliver the secondary beams at the $\pi20$ beamline. 
They are also grateful to the members of the Data Acquisition and Infrastructure Division of RCNP 
and the SPADI alliance for their technical support in implementing the streaming readout data acquisition system.

\let\doi\relax


\begin{thebibliography}{99}
%
\bibitem{PDG} F.~Takahashi et al. (Particle Data Group), to be published in Int. J. Mod. Phys. A 41, 2630011 (2026).
\bibitem{t106} K.~Shirotori et al., KEK/J-PARC-PAC 2025-39 (J-PARC T106 Proposal) (2025) (available at: https://j-parc.jp/researcher/Hadron/en/pac\_2501/pdf/P106\_2025-1.pdf). 
\bibitem{t106proc}  H.~Noumi et al., PoS(HADRON2025)142 (2026).
\bibitem{HDEF} K.~H.~Tanaka et al., Nucl. Phys. A 835, 81 (2010).
\bibitem{e50} H.~Noumi et al., KEK/J-PARC-PAC 2012-19 (J-PARC E50 Proposal) (2012) (available at: http://www.j-parc.jp/researcher/Hadron/en/pac\_1301/pdf/P50\_2012-19.pdf).
\bibitem{HDEFex} K.~Aoki et al.,''Extension of the J-PARC Hadron Experimental Facility: Third White Paper,'' arXiv:2110.04462 [nucl-ex].
\bibitem{COMPASS} P.~Abbon et al. [COMPASS Collaboration], Nucl. Instrum. Meth. A 577, 455 (2007).
\bibitem{AMBER} B.~Adams et al. [COMPASS++/AMBER Collaboration], ``Letter of Intent (Draft 2.0): A New QCD Facility at the M2 beam line of the CERN SPS (COMPASS++/AMBER),'' arXiv:1808.00848 [hep-ex].
\bibitem{brich} N.~Tomida et al., JPS Conf. Proc. 45, 011204 (2026).
\bibitem{t106proc2} T.~Akaishi et al., JPS Conf. Proc. 45, 011110 (2026).
\bibitem{t103} K.~Shirotori et al., JPS Conf. Proc. 45, 011191 (2026).
\bibitem{SRODAQ} K.~Shirotori et al., submitted to Prog. Theor. Exp. Phys. (2026).
\bibitem{mikumari} R.~Honda, IEEE Transactions on Nuclear Science, Vol 70-6, 1102-1109 (2023).
\bibitem{hr-tdc} R.~Honda, M.~Ikeno, C.~S.~Lin, and M.~ Shoji, in Proc. Conf. Rec. 24th IEEE Real Time Conf., Jan. 2020, pp. 1-7, Accessed: May 1, 2024
\bibitem{cirasame} CIRASAME, [online] Available: https://spadi-alliance.github.io/ug-cirasame/
\bibitem{nestdaq} T.~N.~Takahashi, R.~Honda, Y.~Igarashi, H.~Sendai, IEEE Transactions on Nuclear Science, Vol 70-6, 922-927 (2023).
\bibitem{geant4}  J.~Allison et al., Nucl. Instrum. Meth. A 835 (2016) 186-225.
\bibitem{t112}  K.~Shirotori et al., KEK/J-PARC-PAC 2025-40 (J-PARC T112 Proposal) (2025) (available at: https://j-parc.jp/researcher/Hadron/en/pac\_2507/pdf/P112\_2025-16.pdf).
\bibitem{t113}  H.~Noumi et al., KEK/J-PARC-PAC 2025-40 (J-PARC T113 Proposal) (2025) (available at: https://j-parc.jp/researcher/Hadron/en/pac\_2507/pdf/P113\_2025-17.pdf).
%
\end{thebibliography}

\end{document}